\documentclass[useAMS,usenatbib,referee]{mn2e}
\usepackage{epsfig}
\usepackage{graphicx}
\title[Particle acceleration and Non-thermal emissions in  
pulsar magnetosphere]{PARTICLE ACCELERATION AND NON-THERMAL EMISSION
IN PULSAR OUTER MAGNETOSPHERIC GAP}
\author[J. Takata, H.-K. Chang and S. Shibata]{J. Takata,$^{1}$\thanks{E-mail:
takata@tiara.sinica.edu.tw}, H.-K. Chang$^{2}$, and S. Shibata$^{3}$\\
$^{1}$Institute of Astronomy and Astrophysics,  and
 Theoretical Institute for Advanced Research in Astrophysics,
Academia Sinica; and National Tsing Hua University,
Taipei Taiwan\\
$^{2}$Department of Physics and Institute of Astronomy, National
Tsing Hua University, Hsinchu, Taiwan \\
$^{3}$Department of Physics, Yamagata University, Yamagata,
Japan 
} 
\begin{document}

\date{}

\pagerange{\pageref{firstpage}--\pageref{lastpage}} \pubyear{2002}

\maketitle

\label{firstpage}

\begin{abstract}
A two-dimensional electrodynamic model is used to study particle
acceleration and non-thermal emission mechanisms in the pulsar magnetospheres.
 We solve distribution of the
accelerating electric field with  the emission process and 
 the pair-creation process in meridional plane, which includes the
rotational axis and the magnetic axis. By solving the evolutions of 
the Lorentz factor,  and of the pitch angle, 
we calculate  spectrum in optical through $\gamma$-ray bands with 
the curvature radiation, synchrotron radiation,   and inverse~-Compton
process not only for  outgoing particles, but also for ingoing
particles, which were ignored in  previous studies. 

We apply the theory to the Vela pulsar. We  
 find that the curvature radiation from the outgoing particles is the
major emission process above 10~MeV bands. In soft $\gamma$-ray to hard  
X-ray  bands, the synchrotron radiation from the ingoing primary particles
in the gap dominates in the spectrum. Below hard X-ray bands, the
synchrotron emissions from both outgoing and ingoing particles contribute to
the calculated spectrum. The calculated spectrum is consistent
with the observed phase-averaged spectrum of the Vela pulsar. 

Taking into account the predicted dependency of the emission process and the
emitting particles on the energy bands,  
we  compute the  expected pulse profile in X-ray and $\gamma$-ray bands 
 with a three-dimensional geometrical model. 
We show that  the observed five-peak pulse profile
 in the X-ray bands of  the Vela pulsar is reproduced by the 
 inward and  outward emissions,  and the observed double-peak pulse
profile in $\gamma$-ray bands is explained by the outward emissions. 

We also apply the theory to PSR B1706-44 and PSR B1951+32, for which 
X-ray emission properties
have not been constrained observationally very well, to  predict the
spectral features with  the present outer gap model.  
\end{abstract}

\begin{keywords}
pulsars: general -- radiation mechanisms:non-thermal
\end{keywords}

\section{Introduction}
The Compton Gamma-Ray Observatory (CGRO) had  measured puled $\gamma$-ray
emissions from  younger pulsars (Thompson 2004). 
The multi-wavelength  observations of the $\gamma$-ray pulsars 
have shown that the spectra of the non-thermal emissions 
extend in $\gamma$-ray through  optical bands.  
The observations have also revealed the 
pulse profiles in optical through  $\gamma$-ray bands,  and the polarization
properties of the pulsed optical emissions from the $\gamma$-ray
pulsars (Kanbach et al. 2005; Mignani et al 2007).  
In the future, furthermore, 
the polarization of the X-rays and $\gamma$-rays from the pulsars  
will probably be able to be  measured by ongoing projects (Kamae et
al. 2007; Chang et al. 2007).  
The  multi-wave length observations on the  spectrum, the pulse profile
 and the polarization allow us to  perform a comprehensive 
theoretical discussion for the mechanisms of the 
particle acceleration and of the 
non-thermal emission in the pulsar magnetospheres.

The particle acceleration  and the non-thermal emission processes
in the pulsar magnetosphere have  been mainly  
argued with the polar cap accelerator model 
(Sturrock 1971; Ruderman \& Sutherland 1975) and the outer gap accelerator 
model (Cheng, Ho \& Ruderman 1986a, 1986b). Also, 
the slot gap model (Muslimov and
Harding 2004), which is 
an extension model of the polar cap model,  was  proposed.  
All models predict an acceleration of the particles by an  electric field 
parallel to the magnetic field. In the pulsar magnetosphere, 
the accelerating electric field arises in a charge depletion region 
from the so called Goldreich-Julian charge density (Goldreich \& Julian 1969),
 and the  strong acceleration region in the magnetosphere  depends on th
e model. The polar cap and the slot gap models predict  
a strong acceleration within several stellar radii on the polar cap, 
 and the outer gap model predicts a strong acceleration beyond 
the null charge surface, on which the Goldreich-Julian charge 
density becomes zero.   

As CGRO had observed,  most of the $\gamma$-ray pulsars have 
a double peak structure in the pulse profile in $\gamma$-ray bands.
The outer gap model has been successful in explaining the observed double peak 
structure  (Romani \& Yadigaroglu 1995). 
With the outer gap model, the double peak structure 
is naturally produced as an effect of the aberration and the time 
delay  of the emitted photons by the outgoing  
particles that move toward the light cylinder, which is defined
 by the positions, on which
 the rotating speed with the star is equal to speed of the light.
The axial distance of the light cylinder 
 is $R_{lc}=c/\Omega$, where $c$ is the speed of the light,  
and $\Omega$ is the angular velocity of the neutron star. 

Because the Crab pulsar has the pulse profiles with the double peak
structure in whole energy bands, the outer gap model  can
explain the pulse profiles of the Crab pulsar 
 in optical through $\gamma$-ray bands (Takata \& Chang 2007). 
Furthermore, the outer gap model 
can explain the observed  spectra, and the polarization 
characteristics in optical bands for the Crab pulsar (Cheng et
al. 2000; Hirotani 2007; Jia, et al 2007;  
Takata et al, 2007; Takata \& Chang 2007; Tang et al. 2007). 
Therefore, the outer gap accelerator model has successfully 
explained the results of the multi-wave length observations for the
Crab pulsar.

For the Vela pulsar, multi-peak structure in pulse profiles in X-ray, UV 
and optical bands have been revealed (Harding et al. 2002; Romani et
al. 2005), although the pulse profile in $\gamma$-ray 
bands has two peaks in a single period (Fierro et al. 1998). For example,
 Harding et al. (2002) analysed 
RXTE data and revealed  the pulse profile, which  has at least  
 five peaks in  a single period. 
Because the  two peaks in the five peaks  
are in phase with  the two peaks in
the $\gamma$-ray bands, the two peaks  will be explained by the 
traditional outer gap  model with 
 the outward emissions from the outer gap. However, because other three peaks 
are not expected by the traditional outer gap model,  the origin of 
 the three peaks has not been understood, so far.  

The inward emissions emitted by particles accelerated toward stellar surface 
 will give  one possibility to explain the unexpected three 
peaks of pulse profile in the X-ray bands of the Vela pulsar.  
Because the primary particles are produced near the 
inner boundary of the gap, the ingoing particles feel a small 
part of  whole potential drop in the gap 
before  escaping the gap from the inner boundary, 
although the outward moving particles can feel 
whole potential drop before escaping the gap from the outer boundary.
In previous studies, therefore, the contribution of the 
inward emissions on the spectral calculation 
has been ignored by assuming that its flux is much smaller than that of 
the outward emissions.  
However, it will be true only in the $\gamma$-ray bands with 
the curvature radiation. Because 
main emission mechanism in the X-ray bands will be 
the synchrotron radiation, 
the inward synchrotron emissions with a stronger magnetic field will 
be efficient enough to contribute on the observed emissions below $\gamma$-ray
 bands.  Although  the inward emissions have been consistently dealt 
in the electrodynamic studies (Takata et al. 2006;
Hirotani, 2007), 
only the outward emissions were taken into account for computing the spectrum. 

In this paper, we calculate the spectrum  in optical through $\gamma$-ray 
bands  by taking into account both  inward and outward 
 emissions. Specifically, we solve the electrodynamics in the 
outer gap accelerator
in the two-dimensional plane, which includes the rotational axis and
 the magnetic axis.  We calculate the spectrum of 
 the curvature radiation, synchrotron radiation, 
 and  inverse~-Compton process for the primary and the secondary particles. 
We also discuss the expected pulse profiles from optical through $\gamma$-ray 
bands with a  three-dimensional model for comparison with the observations. 
Furthermore, we apply the
theory to PSRs B1706-44 and B1951+32 to calculate the spectrum,
because the properties of optical and X-ray emissions from the two pulsars
 have  not been understood well.

In section~2, we describe our two-dimensional electrodynamic model following 
Takata et al. (2004, 2006). In section~3, we apply  the theory to 
 the Vela pulsar,  and we discuss the  spectrum and the pulse profile 
in optical through $\gamma$-ray bands. We also show the expected spectra 
of  PSRs B1706-44 and B1951+32. 

\section{Two-dimensional electrodynamic model}
We consider a stationary structure in the meridional plane, which includes 
the magnetic axis and the rotation axis. In the meridional plane, we solve 
the Poisson equation of the accelerating electric field, 
the continuity equations for electron and for positron on 
each magnetic field line, and the  pair-creation process by  
 $\gamma$-ray and the surface X-rays (section~\ref{gapst}).
 We assume that new born particles 
via pair-creation process in the gap is quickly saturated between the 
accelerating force and the curvature radiation back reaction force,
 instead solving the evolutions of the Lorentz factor and of the
 pitch angle of the  particles. To obtain the electric structure, 
the saturation treatment is a good assumption, and  simplifies
 the problem to obtain a 
outer gap structure with an iterating method (section~\ref{boundary}).  
 
In the  method described above, however, we can not calculate
 the synchrotron radiation of the 
new born pairs because we do not solve the pitch angle evolution,  
although the synchrotron radiation is an important emission mechanism 
in the X-ray regions. 
With obtained  electric structure with our dynamic model, therefore, we 
solve the evolutions of the Lorentz factor and of the pitch angle 
of the particles for calculating the synchrotron radiation process 
in the gap (section~\ref{emipro}). In this paper, we take into account 
the curvature radiation, the 
synchrotron radiation, and the inverse~-Compton processes
(section~\ref{spectra}). 
 We consider 
the magnetized rotator that 
 an inclination angle $\alpha$ 
between the rotation axis and 
the magnetic axis  is smaller than $90^{\circ}$. We use 
a dipole magnetic field to solve the structure of the outer gap. 

\subsection{Stationary gap structure}
\label{gapst}
The stationary electric potential, $\Phi_{nco}$, 
 for the accelerating field  is obtained from  (Mestel 1999)
\begin{equation}
\triangle\Phi_{nco}(\mathbf{r})=-4\pi[\rho(\mathbf{r})
-\rho_{GJ}(\mathbf{r})],
\label{poisson}
\end{equation}
where $\rho(\mathbf{r})$ is the space charge density, and
$\rho_{GJ}(\mathbf{r})$ is the Goldreich-Julian charge density 
(Goldreich \& Julian 1969), $\rho_{GJ}=\Omega B_z/2\pi c$. 
We assume that the gap dimension in
the azimuthal direction is much larger than that in the meridional
plane. Neglecting  variation in the azimuthal direction,
we rewrite  equation (\ref{poisson}) as
\begin{equation}
\triangle_{r,\theta}\Phi_{nco}(\mathbf{r})
=-4\pi[\rho(\mathbf{r})-\rho_{GJ}(\mathbf{r})],
\label{basic1}
\end{equation}
 where $\triangle_{r,\theta}$ represents ($r,\theta$)-parts of
 the Laplacian.

The continuity equations for the particles is written as 
\begin{equation}
 \mathbf{B}\cdot\nabla\left(\frac{v_{||}N_{\pm}(\mathbf{r})}{B}\right)=\pm
S(\mathbf{r}),
\label{basic2}
\end{equation}
where $v_{||}\sim c$ is the velocity along the field line, 
  $S(\mathbf{r})$ is the source term due to the pair-creation process, 
 and  $N_+$ and $N_-$  denote
the number density of the outgoing and ingoing   particles
 (i.e. the positrons and electrons in the present case), respectively. 
In the outer gap accelerator, $\gamma+X\rightarrow e^++e^-$  
process is 
the main pair-creation process,  and contributes to the source term
 $S(\mathbf{r})$ in equation (\ref{basic2}). 
The pair-creation rate is calculated  from 
\[
\eta_p(\mathbf{r},E_{\gamma})
=(1-\cos\theta_{X\gamma})c\int_{E_{th}}^{\infty}dE_{X}\frac{dN_X}
{dE_X}(\mathbf{r},E_{X})\sigma_p(E_{\gamma},E_{X}),
\]
where $dE_{X}\cdot dN_X/dE_{X}$ is the X-ray number
density between energies $E_{X}$ and
$E_X+dE_X$, $\theta_{X\gamma}$ is the collision angle
between an X-ray photon and a $\gamma$-ray photon,
$E_{th}=2(m_ec^2)^2/(1-\cos\theta_{X\gamma})E_{\gamma}$ is the
threshold X-ray energy for the  pair creation,  
and  $\sigma_p$ is the pair creation cross-section, which is given by
\begin{equation}
\sigma_{p}(E_{\gamma},E_{X})=\frac{3}{16}
\sigma_{T}(1-v^2)\left[(3-v^4)\ln\frac{1+v}{1-v}-2v(2-v^2)\right],
\label{cross}
\end{equation}
where
\[
v(E_{\gamma},E_{X})=\sqrt{1-\frac{2}{1-\cos\theta_{X\gamma}}\frac{(m_ec^2)^2}
{E_{\gamma}E_{X}}},
\]
and  $\sigma_{T}$ is the Thomson cross section.  In the gap, the GeV photons 
collide with the thermal  X-ray photons from the stellar surface. 
 At the radial distance $r$ from the centre of the star,
 the thermal photon number density between energy
$E_{X}$ and $E_{X}+dE_{X}$ is given by
\begin{equation}
\frac{dN_X}{dE_X}=2\pi\left(\frac{1}{ch}\right)^3
\left(\frac{R_{eff}}{r}\right)^2
\frac{E_X^2}{\exp(E_X/kT_s)-1},
\label{soft}
\end{equation}
where $R_{eff}$ is the effective radius of the emitting region, and
$kT_s$ refers to the surface temperature. For  the values of $R_{eff}$ and
$T_s$, the  observed ones are used. Because the $\gamma$-ray photons 
are emitted in direction of velocity of the particles, 
the collision angle is estimated 
from $\cos\theta_{X\gamma}=v_{||}B_r/cB$, where $B_r$ is the radial
 component of the magnetic field.

To calculate  the source term $S(\mathbf{r})$ in equation (\ref{basic1}) at 
each point,
 we simulate the
photon-photon pair-creation process with Monte Carlo method as
described in Takata et al (2006). Specifically, 
 a $\gamma$-ray may convert into a pair at the distance $s$ with the
 probability that 
\begin{equation}
P_p(s)=\frac{\int^s_01/l_p ds}{l_p},
\label{prob}
\end{equation}
where $l_p=c/\eta_p$ is the mean-free path of the pair-creation. We  determine 
the pair-creation position following the probability of 
 equation (\ref{prob}). 

Most of GeV photons are emitted via the curvature process of the electrons 
and the positrons with the Lorentz factor of $\sim 10^{7.5}$. 
 The power per unit energy of the curvature process is written as 
\begin{equation}  
P_c(R_c,\Gamma,E_{\gamma})=  
\frac{\sqrt{3}e^2\Gamma}{hR_c} 
F(x),  
\label{curate} 
\end{equation}   
where $x\equiv E_{\gamma}/E_c$,  
\begin{equation}  
E_c=\frac{3}{4\pi}\frac{hc\Gamma^3}{R_c},  
\label{critic} 
\end{equation}  
and  
\begin{equation}  
F(x)=x\int_x^{\infty}K_{5/3}(t)dt,  
\end{equation}  
where $R_{c}$ is the curvature radius of the magnetic field line,
$\Gamma$  is the Lorentz factor of the particles,
 $K_{5/3}$ is the modified Bessel function of the  order 5/3, 
$h$ is the Planck constant, and $E_{c}$   
gives the characteristic curvature photon energy. 

We assume the saturated motion to obtain the 
stationary electric field structure. By assuming that the particle's 
motion immediately saturates in the    
balance between the electric and the radiation reaction forces, the Lorentz 
factor at each point is calculated from 
\begin{equation}   
\Gamma_{sat}(R_c,E_{||})=\left(\frac{3R_c^2}{2e}E_{||}+1\right)^{1/4}.
\label{gamma}   
\end{equation} 
\subsection{Boundary conditions}
\label{boundary}
To solve the Poisson equation (\ref{basic1}), we impose
 the boundary conditions on the four boundaries, which are called 
as inner, outer, upper and lower boundaries.  
The lower and upper boundaries are laid on the magnetic 
surfaces, and the lower boundary is defined by the last~-open field line. 
The inner and the outer boundaries are defined by the surfaces on which 
the accelerating electric field is vanishes, that is,  $E_{||}=0$. 

We anticipate that the inner, upper, and lower boundaries are 
directly linked with the star without the potential drop. We then impose that 
the accelerating potential is equal to zero, that is $\Phi=0$, 
 on the inner, upper and lower boundaries. We note that the position of the 
inner boundary is not free because 
the Dirichlet- and Neumann-type conditions are imposed on  it. By moving the 
inner boundary step by step iteratively, we seek the boundary that
 satisfies the required conditions.  

As demonstrated in Takata et al. (2004), 
the inner boundary, which is satisfied 
the required conditions, comes to the position, on which the condition 
that $j_g+j_2-j_1=B_z/B$ is satisfied, where $j_g$ is 
 the current in units of the Goldreich-Julian value 
 carried  by the pairs produced in  the gap,  $j_1$ is  
 the current carried by the positrons coming into the gap through the 
inner boundary and  $j_2$ is  
 the current carried by the electrons coming into the gap through the 
outer boundary. We find that if there is a current in the outer gap, 
the inner boundary is shifted  from the null charge surface, where is the 
inner boundary for the vacuum gap, $j_g=j_1=j_2=0$. 
Total current in unit of the Goldreich-Julian charge density 
is described by $j_{tot}=j_g+j_1+j_2$ and is constant along a field line. 
The model parameters are the current $(j_1, j_2, j_g)$ and
 the inclination angle $\alpha$. A more detail discussion on 
the boundary conditions and the model parameters  is  seen 
in Takata el al. (2004,2006). 

\subsection{Particle motion}
\label{emipro}
To compute the synchrotron radiation with the pitch angle, 
we use the electric field distribution in the outer gap obtained by 
 the method described in section~\ref{gapst} to 
solve the equation of motion, which describes 
the evolutions of the pitch angle of the  particle.
For the Vela pulsar, the inverse~-Compton process is less significant 
 for energy loss of the particles than the synchrotron and curvature 
radiation. 
The momenta of the parallel ($P_{||}/m_e c=\sqrt{\Gamma^2-1}\cos\theta_{p}$) 
and perpendicular ($P_{\perp}/m_e c=\sqrt{\Gamma^2-1}\sin\theta_{p}$) 
to the magnetic field are, 
respectively, described as (Harding et al. 2005; Hirotani 2006)
\begin{equation}
\frac{dP_{||}}{dt}=eE_{||}-P_{sc}\cos\theta_p,
\label{paraeq}
\end{equation}
and \begin{equation}
\frac{dP_{\perp}}{dt}=-P_{sc}\sin\theta_p+\frac{c}{2B}\frac{dB}{ds}P_{\perp},
\label{perpeq}
\end{equation}
where $\theta_p$ is the pitch angle, $P_{sc}$ represents the radiation drag 
of the synchrotron and curvature radiation, and the second term on the right
 hand side on equation~(\ref{perpeq}) represents the adiabatic change 
along the dipole field line. The radiation drag, $P_{sc}$, of 
the synchrotron-curvature radiation is described by (Cheng \& Zhang 1996), 
\begin{equation}
P_{sc}=\frac{e^2c\Gamma^4Q_2}{12r_c}\left(1+\frac{7}{r_c^2Q_2^2}\right),
\end{equation}
where 
\begin{equation}
r_c=\frac{c^2}{(r_B+R_c)(c\cos\theta_p/R_c)^2+r_B\omega_B^2},
\end{equation}
\begin{equation}
Q^2_2=\frac{1}{r_B}\left(\frac{r_B^2+R_cr_B-3R_c^2}{R_c^3}\cos^4\theta_p
+\frac{3}{R_c}\cos^2\theta_p+\frac{1}{r_B}\sin^4\theta_p\right)
\end{equation}
\begin{equation}
r_B=\frac{\Gamma m_ec^2\sin\theta_p}{eB},~~~~\omega_B=\frac{eB}{\Gamma m_ec}.
\end{equation}
We solve the equations of the motion, (\ref{paraeq}) and (\ref{perpeq}), 
 up to the light cylinder for the outgoing
 particles (positrons) and to the stellar surface 
for the ingoing particles (electrons). 
 The initial pitch angle of the new born pairs  is determined
  by the angle between the propagating 
direction of  $\gamma$-rays and the direction of the magnetic 
field at the pair-creation point. 

\subsection{Spectra of radiation processes}
\label{spectra}
We calculate the curvature radiation, synchrotron radiation and 
 inverse~-Compton process. 
The power of the synchrotron-curvature process for 
 a particle  is calculated from (Cheng \& Zhang 1996),
\begin{eqnarray}
\frac{dP}{d E}&=&\frac{\sqrt{3}e^2\Gamma E}{4\pi r_c E_c}
\Bigg\{\left[\int_{\omega/\omega_c}^{\infty}K_{5/2}(y)dy
-K_{2/3}(\omega/\omega_c)\right] \nonumber \\
&+&\frac{[(r_B+R_c)(c\cos\theta_p/R_c)^2
+r_B\omega^2_B}{c^4Q_2^2}\left[\int_{\omega/\omega_c}^{\infty}K_{5/2}(y)dy
+K_{2/3}(E/E_c)\right]\Bigg\}. 
\end{eqnarray}

Very high-energy $\gamma$-rays are produced by the inverse~-Compton process. 
The emissivity of the inverse~-Compton process for a particle 
is calculated from 
(Takata \& Chang 2007)
\begin{equation}
\frac{d P_{in}}{dEd\Omega}=
 D^2 
\int_0^{\theta_c} (1-\beta\cos\theta_0) 
I_b/h \frac{d\sigma'}{d\Omega'}d\Omega_0, 
\end{equation}
where $\theta_0$ is the angle between the directions of the particle motion 
and the propagating direction of the background photon, 
$D=\Gamma^{-1}(1-\beta\cos\theta_1)^{-1}$ with 
$\theta_1$ being 
 the angle between the directions of the particle motion and 
the propagating direction of  the scattered photons.  
The angle $\theta_c=\sin^{-1}R_{*}/r$,  with $R_*$ being stellar
surface, 
expresses the size of the star seen from the point $r$, 
$d\Omega_0$ is the solid angle of the background radiation. 
 The differential cross section is given by the Klein-Nishina formula 
\begin{equation}
 \frac{d\sigma'}{d\Omega'}=\frac{3\sigma_T}{16\pi}
\left(\frac{\epsilon'_1}{\epsilon'_0}\right)^2
\left(\frac{\epsilon'_1}{\epsilon'_0}+\frac{\epsilon'_0}{\epsilon
_1}-\sin^2w'
\right),
\end{equation}
where $\epsilon'_0$ and $\epsilon'_1$ are the energy of the background 
and the scattered photons in units of the electron rest mass energy, 
respectively, in the electron rest frame, and  they 
 are connected each other by $\epsilon'_1=\epsilon'_0/[1+\epsilon'_0
(1-\cos w'_s)]$. The scattering angle $w'$ is defined by 
\begin{equation}
\cos w'=\sin\theta'_0\sin\theta'_1\cos(\phi'_0-\phi'_1)
+\cos\theta'_0\cos\theta'_1
\end{equation}
where $\theta'_0$ and $\theta'_1$  
 are the polar angle  of the propagating 
direction of the background and the scattered photons, respectively, 
 measured from the particle motion in the electron rest frame, 
 $\phi'_0$ and $\phi'_1$ are the azimuthal directions of the photons. 

In  the present model, we adopt the surface thermal emission as 
the background field of the inverse~-Compton process. 
In such a case, the intensity  $I_b$ is 
described by the Planck distribution that 
\begin{equation}
I_b(E_X)=\frac{2E_X^3/h^2c^2}{\exp(E_X/kT_s)-1}.
\end{equation}

We calculate the emissions from  the following 
four  different kind of electron and positron, 
\begin{enumerate}
\item primary particles, which are created inside the
gap,  and are accelerated up to ultra-relativistic energy;
\item secondary particles created outside the gap via the
photon-photon pair-creation process by the surface thermal 
 X-ray photons;

\item secondary particles created outside the gap via the
photon-photon pair-creation process by  
the magnetospheric non-thermal  X-ray photons;

\item secondary particles created outside the gap  via  the magnetic
pair-creation.  
\end{enumerate}
For the magnetic pair-creation, we assume that 
 $\gamma$-ray photons are converted into the  pairs at 
the point, on which  the  condition that 
$E_{\gamma} B\sin\theta_p/B_{cr}=0.2m_ec^2$ (Muslimov \& Harding 2003)  
is satisfied, where  $B_{cr}=4.4\times 10^{13}$~Gauss
 is the strength of the critical magnetic field. 
The TeV photons emitted by the inverse~-Compton process also produce 
the secondary  pairs. Because the flux of the 
TeV radiation is much smaller than the flux of GeV radiation 
(Figure~\ref{totspe}), we ignore the emissions for the pairs produced by the 
TeV photons.

\section{Results}
\begin{figure}
\begin{center}
\includegraphics[width=14cm, height=7cm]{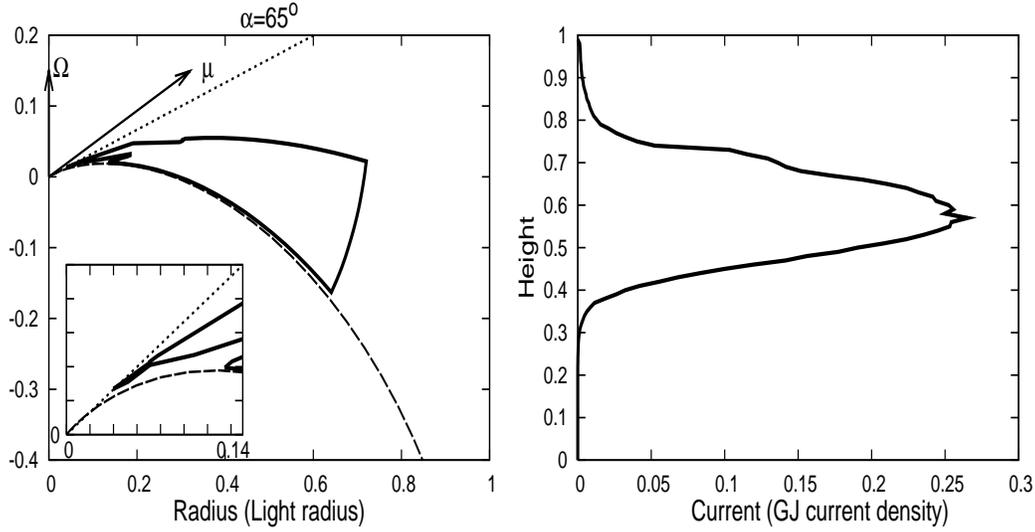}  
\caption{Left; The geometry of the outer gap accelerator. The solid
line is the boundary of the gap. The dashed line  show the
last~-open line. For the dotted line, the condition that $B_z=0.26B$ is
satisfied. Right; The trans-field structure of the current. 
The insinuation angle is $\alpha=65^{\circ}$. }
\label{gapstr}
\end{center}
\end{figure}

\begin{figure}
\begin{center}
\includegraphics[width=16cm, height=7cm]{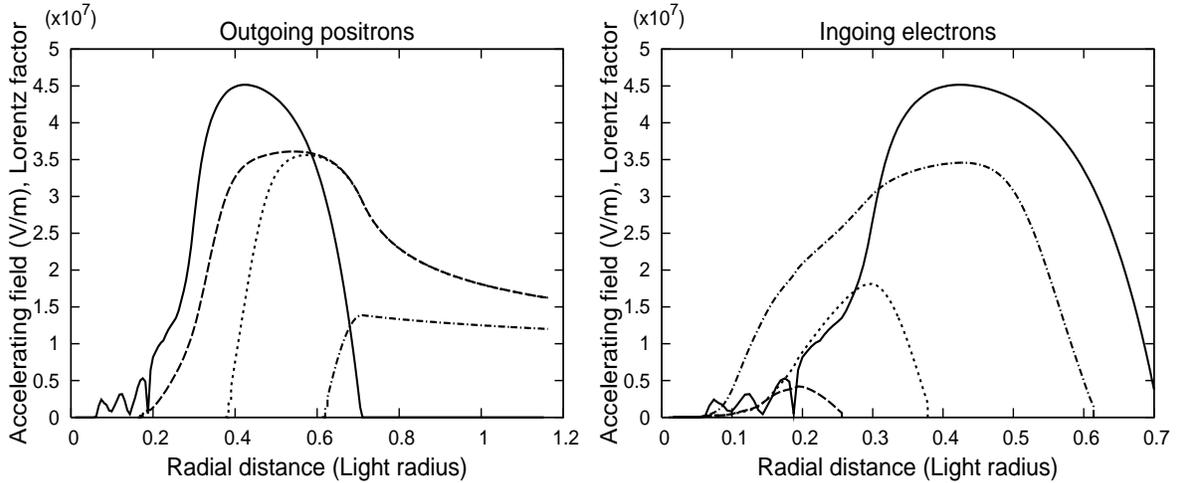}  
\caption{Evolution of the Lorentz factor of the outgoing positrons 
(the left panel) and the ingoing electrons (the right panel) along a magnetic
field line.  The dashed,
dotted and dashed-dotted line are for the particles produced near the 
inner boundary, produced the middle of the gap and produced near the
outer boundary, respectively. The solid lines show the distribution 
of the accelerating electric field along the field line.}
\label{motion}
\end{center}
\end{figure}

\subsection{The Vela pulsar} 
In this section, we apply the theory to the Vela pulsar. 
With the outer gap model, 
 Romani \& Yadigaroglu (1995) used $\alpha\sim 65^{\circ}$ 
to produce the pulse profile of the $\gamma$-ray emissions from the Vela 
pulsar. This inclination angle will be consistent with the inclination
 angle $\alpha\sim 60^{\circ}$ for fitting  the polarization  swing 
of the radio emissions (Krishnamohan \& Downs 1983). 
In this paper, we adopt $\alpha=65^{\circ}$ to 
the Vela pulsar. For the thermal emissions from the stellar surface, 
we use the temperature $kT_s=0.1$~keV with the effective radius 
$R_{eff}=4.4 (d/0.25~\mathrm{kpc})^2$~km (Manzali et al. 2007).  
We assume a  small injection 
of the particles through the boundaries with $j_1=10^{-5}$ and $j_2=0$. In such
a case, the total current is carried by the pairs created inside 
the gap, that is, $j_{tot}\sim j_{gap}$. 
In this paper, we show the outer gap structure that extends
 from near the stellar surface.   

\subsubsection{Outer-gap structure}
Figures~\ref{gapstr} 
 summarises the outer gap geometry in the magnetosphere
 and the current in the gap. 
Figure~\ref{gapstr} 
(the thick solid line in the left panel)
 shows the boundary  
of the outer gap. In side of the thick solid line, the particles are
 accelerated along the magnetic field lines. The
dashed line in the figure shows the last~-open field line. The
trans-field distribution of the current is shown in Figure~\ref{gapstr} 
(right). The abscissa refers the current in unit of the
Goldreich-Julian value, $\Omega B/2\pi $, and the vertical line represents
the height measured from the last~-open field line. For example, we can
read from the figure that about 26\% of the Goldreich-Julian current
runs through the gap at the height of 
the 60\% of the thickness measured from 
the last~-open field line.

As discussed in Takata et al. (2004, 2006), the inner boundary with a 
 small particle injections is
 located at the position,  on which $B_z=j_g B$ is satisfied.  
For example, if $j_g=0$, that is, if the gap is vacuum, the inner
boundary comes to the position, on which $B_z=0$, and therefore  
the null charge surface becomes the inner boundary of the gap. 
If there is the pair-creation in the gap, on the other hand, 
the inner boundary is shifted
toward the stellar surface due to the current. 

The sub-panel in the left panel of  Figure~\ref{gapstr} 
zooms the region around
the inner boundary.  The dotted line in the figure shows the 
position, on which the condition $B_z=0.26B$, where 
$j_g=0.26$ is the maximum current in the outer gap , 
is satisfied. We can see that the cusp of the inner
boundary, where the maximum current runs through on the field line, 
 is located on the dotted line. Thus, the outer gap can extend
 near the stellar surface with a large current. 
 
\subsubsection{Particle motion}
Figure~\ref{motion} summarizes 
the distribution of the accelerating electric field 
 and the evolutions of the accelerated particles along the
field line, which  penetrates  the outer gap at the height
 of 50\% of thickness.  The left and the right panels 
show the evolutions of the Lorentz factors 
of the outgoing and ingoing particles, respectively, and represent 
for the three kinds of the 
particles, which are produced near the inner boundary (the dashed line), 
middle of the gap (the dotted line)  
and near the outer boundary (the dashed-dotted line). 
The solid lines in the figure 
show the distribution of the
strength of the accelerating electric field along the field line. 

For example, a positron produced middle point of the gap (Figure~\ref{motion}, 
the dotted line in the left panel) is outwardly  
accelerated by the electric field, 
and is immediately boosted above $10^7$ on the Lorentz factor. This
positron escapes from  the outer gap around $r\sim 0.7$, where 
the accelerating field vanishes. Because there
is no accelerating electric field outside the gap, 
the particles loose their energy
via the curvature radiation. Because the cooling length of the
curvature radiation, $l_{cool}/R_{lc}\sim
1.6(R_c/R_{lc})^2(\Gamma/10^7)^{-3}(\Omega/100\mathrm{s}^{-1})^{-1}$, 
 becomes  comparable with the light radius at 
 the Lorentz factor of $\Gamma\sim 10^7$, the particles escape from the
light cylinder with the Lorentz factor of $\Gamma\sim 10^7$. If the
particle was not accelerated above   $\Gamma\sim 10^7$ inside the
gap such as the positrons produced near the outer boundary
(Figure~\ref{motion}, the dashed-dotted line in the left panel),
 the Lorentz factor of the particles does not
change very much between the outer boundary and the light cylinder. 

An electron produced middle of the gap (
Figure~\ref{motion}, the dotted line in the right panel) 
is inwardly accelerated by the electric field, and are
also immediately boosted above $10^7$ on the Lorentz factor 
in the gap. The accelerated electrons  
significantly loose their energy near and outside the 
inner boundary  due to the
curvature radiation.  This is because the
curvature radius near the stellar surface is about 
$R_c\sim 10^7$~cm,  and therefore the
radiation drag force near the stellar surface is strong  
so that the cooling length is much shorter than light radius, 
if the Lorentz factor is larger $10^7$. 
The primary electrons reach the stellar surface with the 
Lorentz factor of $\Gamma\sim 10^6$.

\subsubsection{Spectrum}
\label{spectrum}
\begin{figure}
\begin{center}
\includegraphics[width=7cm, height=7cm]{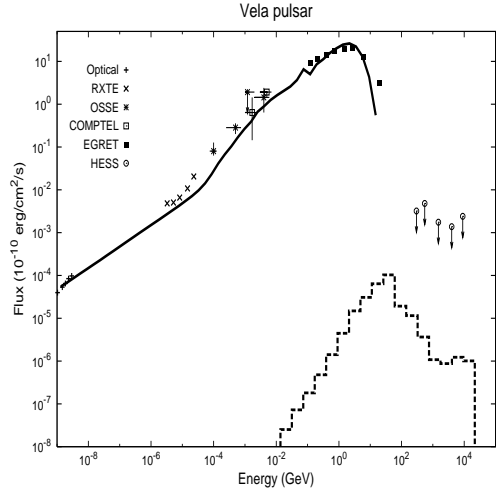}  
\caption{Spectrum of the Vela pulsar. The solid line shows the total
spectrum of the curvature and the synchrotron radiation from the
primary and the secondary particles. The dashed line shows the
spectrum of the inverse~-Compton process of the primary particles. 
The observation data are taken from Shibanov et al. (2003) for
optical, Harding et al (2002) for RXTE, Strickman et al. (1996) for
OSSE and COMPTEL,  Fierro et al. (1999) for EGRET and Konopelko et al. (2005) 
for HESS.}
\label{totspe}
\end{center}
\end{figure}

\begin{figure}
\begin{center}
\includegraphics[width=16cm, height=7cm]{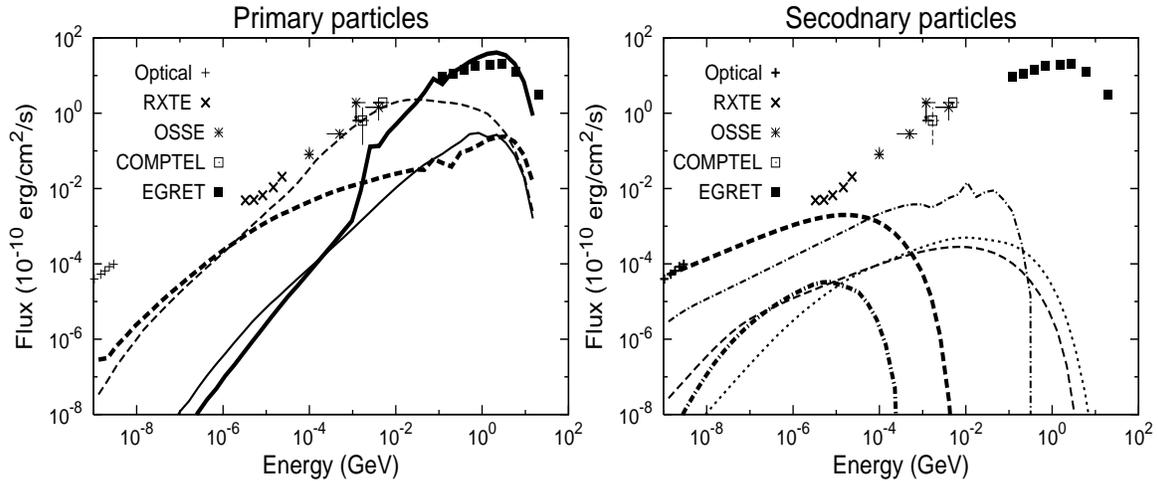}  
\caption{Spectrum of the Vela pulsar. Left; Spectra of the 
the curvature (the solid lines) and synchrotron (the dashed lines) 
radiation from the primary particles. The thick and thin lines show 
the emissions
from the outgoing particles and ingoing particles, respectively.
Right; Spectra of the synchrotron radiation of the secondary
particles. The dashed line show the synchrotron spectra of the  pairs 
produced by the pair-creation process by the magnetospheric
X-rays. The dashed-dotted lines are the spectra of the pairs 
produced by the pair-creation process with surface 
thermal X-rays. The dotted line shows the spectra of the secondary
pairs produced via the  pair-creation process with the strong
magnetic field.   The thick and thin lines show the emissions
from the outgoing particles and ingoing particles, respectively.
 }
\label{spe1}
\end{center}
\end{figure}
Figure~\ref{totspe} shows the calculated spectrum from the optical through TeV 
bands. Figure~\ref{totspe} (the solid line) 
 shows the spectrum of the total emissions, 
which include the curvature and the synchrotron
radiation from the primary and the secondary particles. The dashed
line is the inverse~-Compton spectrum of the primary particles. The
observational data of the phase-averaged spectrum are 
also plotted. We find that  the calculated spectrum is consistent with
the observations in  whole energy bands. 

In Figure~\ref{spe1},  we decompose the total spectrum 
(Figure~\ref{totspe}, the solid line) into the
components of  the primary particles (the left panel) and of the secondary
particles (the right panel). In the left panel, the 
solid lines and the dashed line show the spectra of the
curvature radiation and the synchrotron radiation of the primary particles, 
respectively.  The thick line and the thin line represent 
  the spectra of the outward emissions for  the  outgoing 
positrons and of  the inward emissions for the ingoing electrons, respectively.

For the outward emissions of the primary particles,  
we can see that the ratio of the radiation powers of
the curvature radiation $P_c$ (Figure~\ref{spe1}, the thick solid line
in the left panel) and the synchrotron radiation $P_s$ 
(Figure~\ref{spe1}, the thick dashed line in the left panel)
is about $P_c/P_s\sim10^3$. 
 The perpendicular momentum to the magnetic field lines
quickly decreases via the synchrotron radiation, and its cooling length 
is much shorter than the gap width. Therefore the
synchrotron radiation from the primary particles are efficient only near 
 the pair-creation position, which is around the inner boundary of the gap. 
On the other hand, the outward curvature
radiation of the outgoing particles takes place  at whole outer gap, 
because the particles are always accelerated by the electric field 
in the gap.  In such a case,  
the ratio of the total powers is estimated with   
$P_c/P_s\sim(2e^2\Gamma^4\delta s/R_c^2)/mc^2\Gamma_{\perp}\sim 2\cdot 10^3
 (\Gamma/10^7)^4(\Gamma_{\perp}/10^3)^{-1}(\delta s/0.5R_{lc})$, 
where 
$\delta s$ is the gap width. This estimated value $P_c/P_s$ explains 
the ratio between the calculated fluxes of the outward curvature
emissions  and of the synchrotron emissions.  Although the total power of  
the outward synchrotron radiation is smaller  than 
that of the curvature radiation, 
 the emissions become 
 important below 1~MeV bands as Figure~\ref{spe1} 
(the thick dashed line in the left panel)  shows. 

For ingoing primary electrons (Figure~\ref{spe1},
 the thin lines in the left panel), 
the travel distance in the gap  before escaping from the inner
boundary is much shorter than the gap width, because the most 
pairs are produced near the inner boundary.
And, because the maximum Lorentz factor of the ingoing electrons is 
$\Gamma\sim 10^7$, which is smaller than that of the outgoing positrons, 
the ratio of the radiation powers $P_c/P_s$ becomes about 
unity as  Figure~\ref{spe1} 
(the thin solid and dashed lines in the left panel) shows. 
 
With Figure~\ref{spe1} (the right panel), 
we show the spectra of the synchrotron emissions  for   
three kinds of the secondary pairs; the dashed lines show the spectra for the 
secondary pairs produced by the magnetospheric X-rays, 
the dashed-dotted lines represent the spectra 
for the pairs produced by  
the surface X-rays,  and the dotted line shows the spectrum 
for the pairs produced via the magnetic pair-creation process. 
 The thick and thin dashed lines represent the synchrotron 
spectra from the outgoing and ingoing moving pairs, respectively. 

As Figure~\ref{spe1} (the dashed-dotted lines) shows, the synchrotron emissions
 from the outgoing pairs produced by
the surface X-rays are much fainter than that from the ingoing particles. 
For the outwardly propagating $\gamma$-rays, the pair-creation 
process  by the surface X-ray occur with tail-on-like collision, that is, 
$1-\cos\theta_{X\gamma}\ll 1$. For the ingoing
 propagating $\gamma$-rays, on the other hand, the pair-creation 
process  by the surface X-ray occur with head-on-like collision, that is, 
$1-\cos\theta_{X\gamma}\sim 2$. This difference of the collision angle 
produces a large difference in the mean free path so that  the mean-free path 
of the outgoing $\gamma$-rays is much longer 
than that of the ingoing $\gamma$-rays. Because a smaller number 
of the outgoing secondary pairs than that of the ingoing secondary pairs 
 are produced by the surface X-rays,   
the flux of the synchrotron emissions for the outgoing pairs 
is much fainter than  that for the ingoing pairs. 

 For the secondary pairs produced by the 
magnetospheric X-ray photons, on the other hand, 
 the total energy of synchrotron emissions for the outgoing 
particles is much  larger  than that for the ingoing particles 
(Figure~\ref{spe1}, the dashed lines in the right panel). 
Because the collision angles with the magnetospheric X-ray 
 are not difference 
between the outer outgoing  and ingoing $\gamma$-ray photons, the difference 
of  number of the created pairs originates from  the difference of 
 the number of the $\gamma$-ray photons. 
Because  the outgoing $\gamma$-rays is more than 
that the ingoing $\gamma$-rays as  Figure~\ref{spe1} (the left panel) 
shows, more outgoing secondary pairs is produced  than 
 the ingoing secondary pairs. Therefore, the synchrotron 
emissions for the outgoing secondary pairs produced 
by the magnetospheric X-rays  are brighter than that for the ingoing 
secondary pairs. This will be the  main reason of the  fact that 
the Crab pulsar has the  double peak pulse profile  structure 
 in whole energy bands (section~\ref{crab}).

Using Figure~\ref{spe1}, we know which emission process is important in
 the spectrum in  different energy bands. 
Above 10~MeV, 
we see that the curvature radiation of the outgoing primary particles
 (Figure~\ref{spe1}, the thick solid line in the left panel)
 dominates other  emission processes, and  
explains the EGRET observations,  as  previous studies 
discussed (Romani 1996; Takata et al. 2006). We find that 
between  100~keV and  10~MeV bands, the  synchrotron radiation
 of the ingoing primary particles 
 (Figure~\ref{spe1}, the thin dashed line in the right panel) 
is major emission process.  In soft X-ray bands, 
the synchrotron emissions  for both ingoing and outgoing 
particles,  and of both primary and the secondary particles 
 all contribute to the calculated spectrum to explain RXTE observations, 
as Figure~\ref{spe1} shows.  
 In optical bands,
 the synchrotron emissions for the secondary pairs (Figure~\ref{spe1}, right) 
explain the observations.

This energy dependency of the emission processes in the spectrum of the
 Vela pulsar is quite different from the emission mechanism of the 
Crab pulsar, of which the  synchrotron self inverse~Compton process of the 
outgoing secondary pairs  always dominate other  emission 
processes (section~\ref{crab}). 
Unlike the double-peak pulse profile in whole energy bands 
of the Crab pulsar,  
therefore, we expect that the properties of the 
pulse profile,
 such as the peak position and number of the peaks in a single period, 
depend on the energy bands.

\subsubsection{Pulse profile}
\label{pulsep}

\begin{figure}
\begin{center}
\includegraphics[width=12cm, height=7cm]{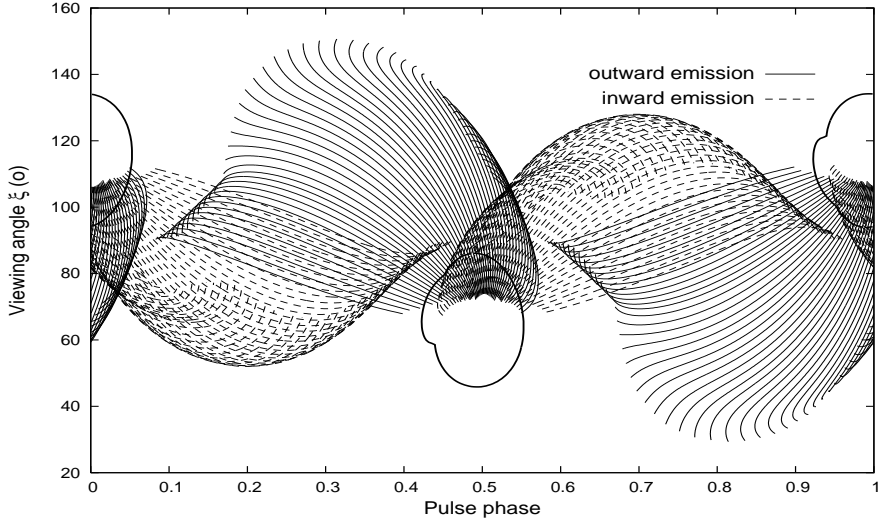}  
\caption{Emission region projected onto $(\xi, \Phi)$-plane for the
magnetic surface, which has a  polar angle of
$\theta_p(\phi)=0.95\theta_l(\phi)$ on the stellar surface. An 
inclination angle is $\alpha=65^{\circ}$ and the emission region
extends 
from $r_{in}=0.5r_n$ to $r=R_{lc}$. and extends around the rotation
axis with the azimuthal of of $180^{\circ}$. The solid  lines and the
 dashed lines corresponding to the outward emissions and the inward
emissions, respectively.  The thick solid circles show the points 
 of the radial distance $r=0.05R_{lc}$ (section~\ref{pulsep})
}
\label{map}
\end{center}
\end{figure}

\begin{figure}
\begin{center}
\includegraphics[width=16cm, height=7cm]{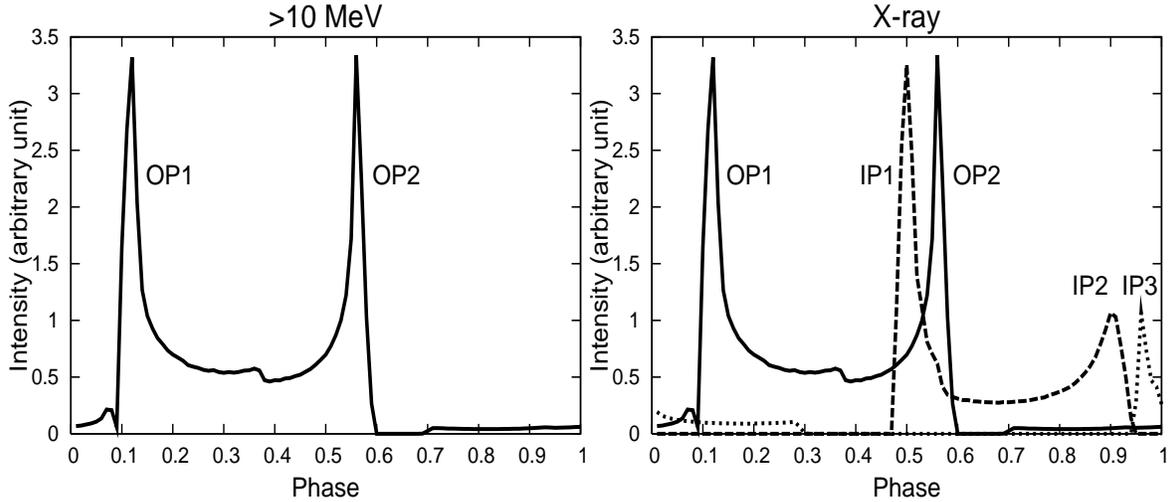}  
\caption{Pulse profile for the Vela pulsar. Left; Expected pulse
profile in $\gamma$-ray bands by the outward emissions. Right;
Expected pulse profile in X-ray bands. The solid line represents the
pulse profile by the outward emissions. The dashed and the dotted line
show the pulse profiles by the inward emissions of outside and inside 
of the null charge surface, respectively. }
\label{pulse}
\end{center}
\end{figure}
In this section, we discuss the expected pulse profiles in $\gamma$-ray 
bands, in soft X-ray bands, and in optical/UV bands 
 with a  three-dimensional model.  
To adopt  a more consistent  three-dimensional 
geometry of  the  emission region, 
we use the solved two-dimensional gap structure
 (Figure~\ref{gapstr}, the left panel) presented in previous sections. 
As Figure~\ref{gapstr} shows,  
 the inner boundary of the  outer gap is located at 
the radial distance $r_{in}\sim 0.5r_{n}$, where $r_n$ is the radial 
distance to the inner boundary,  and is $\sim 0.1R_{lc}$. 
Then, we assume the three-dimensional gap structure that 
the inner boundary on every field line, which penetrates the outer gap, 
 is located at the position satisfied the radial distance 
$r_{in}(\phi)\sim 0.5r_n(\phi)$. 
Because the radial distance to the null charge 
surface varies with the azimuth $\phi$, the radial distance to the 
inner boundary also varies with the azimuth. 
 The lower boundary of the three-dimension gap 
is defined by the surface of the 
last~-open field lines. For the upper boundary, we assume 
the magnetic surface, of which the 
polar angle $\theta_p$ on the polar cap is $\theta_u(\phi)=0.9\theta_{l}
(\phi)$, where $\theta_l$ is the polar angle of the last~-open field lines. 
We assume that the outer gap expands around the rotational axis 
with the azimuthal angle of $180^{\circ}$. 
We note 
that in this paper we discuss the peak positions, 
which do not depend the emissivity 
distribution very well, in the pulse profile 
by ignoring  the distribution of the emissivity. 
Although 
we need a  more detailed model, which deals the three-dimensional 
 distribution of 
 electric field and  of  emissivity,  
 to discuss the phase-resolved spectra with the pulse profile, 
the study of the phase-resolved spectrum is beyond the  scope of 
this work.

We can anticipate that the emission direction in the observer frame 
is coincide with 
the particle motion. For example, the particle motion along the field lines in 
 the north hemisphere, is described by $\mathbf{v}=\pm v_0\mathbf{b}
+v_{co}\mathbf{e}_{\phi}$, where the plus (or minus) sign represents
 the outgoing (or ingoing) particles, $\beta_0$ is the velocity along 
the magnetic field line and is determined by the condition $|\mathbf{v}|=c$, 
$\mathbf{b}$ is the unit vector of the magnetic field, for which 
we adopt a rotating dipole field in the observer 
frame,  $v_{co}$ is 
the co-rotating velocity, and $\mathbf{e}_{\phi}$ is the unit vector of 
the azimuthal direction.  To compute the pulse profile, 
the emission direction, $\mathbf{n}\equiv \mathbf{v}/c$, is 
interpreted in terms of the viewing angle $\xi=n_z$ and 
the pulse phase $\Phi=-\phi_e-\mathbf{n}\cdot \mathbf{r_e}/R_{lc}$, where 
$\phi_e$ and $r_e$ is the azimuthal direction of and the radial distance 
 to the emission point, respectively. 
Figure~\ref{map} shows the emission regions  projected  
 onto $(\xi,\Phi)$-plane, where we adopt 
 an inclination angle of $\alpha=65^{\circ}$. 
We choose the magnetic surface, of   
 which the polar angle of the footpoint on the stellar surface is given
 as $\theta(\phi)=0.95\theta_{l}(\phi)$. 
The solid lines in  Figure~\ref{map} are corresponding to 
 the outward emissions, and the dashed lines are corresponding to 
 the inward emissions. 
We can expect that the inward emissions also make peaks in the pulse profile; 
for example, the peaks due to the inward emissions appear around 
0.5~phase and 0.9~phase for the observer with the viewing angle  of 
$\xi\sim 100^{\circ}$. 

We discuss the expected pulse profile for each energy band. 
According to  Figure~\ref{spe1}, 
 only outward curvature  emissions contribute to 
 the spectrum above 10~MeV.  To calculate the pulse profile in $\gamma$-ray
 bands, therefore, we take into account  the  only  outward emissions, which 
 take place from
the inner boundary to  the light cylinder.   
Figure~\ref{pulse} (the left panel)
 shows the expected pulse profile in $\gamma$-ray bands with the 
viewing angle $\xi\sim 97^{\circ}$. 
Because only outward emissions contribute to the emissions, 
we obtain double peak 
pulse profile, as  previous studies have proposed 
(Romani \& Yadigaroglu 1995; Romani 1996). 

Unlike above 10~MeV,  both outgoing  and ingoing particles contribute 
to the spectrum of  around 1~keV with the synchrotron emissions 
 as Figure~\ref{spe1} shows. 
The right panel of Figure~\ref{pulse} shows the pulse profile, which includes 
both outward and inward emissions. For the inward emissions, we restricted 
 the radial distance of the emission region with  $r_s<r\le3r_n$, where 
the upper boundary $3r_n$ comes from  the results of the dynamic model, 
in which  very few ingoing 
 pairs are produced beyond the radial distance $r>3r_n$. 
We expected that the pulse positions 
 of the outward synchrotron 
emissions are aligned with the phase of 
the peak above 10~MeV (OP1 and OP2 in  Figure~\ref{pulse}). 
The dashed line and the dotted line represent 
 the pulse positions of the inward 
emissions beyond and below the null charge surface, respectively.
 As the dashed line shows,
 the inward emissions beyond the null charge surface make another 
two peaks, which are denoted with IP1 and IP2 in the figure, and the 
inward emissions below the null charge surface 
make one  peak, which is denoted with IP3.
Therefore, the  present model predicts a multi-peak structure of 
the pulse profile in X-ray bands for the Vela pulsar.

We compare the expected peak phases  in X-ray bands with the 
observed peak phases  by 
RXTE (Harding et al. 2002), which indicates 
five peaks in a single period. In RXTE observations, the phase of 
the two peaks, which are denoted Pk~1 and Pk~2-soft in figure~1 in Harding 
et al. (2002), are aligned with the phases of the two peaks 
in the $\gamma$-ray 
bands. With the present model, these two components of RXTE 
are explained by the synchrotron radiation of the outgoing particles, which 
produce two peaks (OP1 and OP2) relating  
with the $\gamma$-ray pulse profiles. 
The observed other two peaks, which are  denoted  Pk~2-hard and Pk~3 
 in figure~1 in Harding et al. (2002), are corresponding to 
the model peaks denoted with IP1 and IP2, which come from the synchrotron 
emissions of the ingoing particles beyond the null surface. 
Remarkably, the  present model produces  the observed 
phase-separation ($\sim 0.4$) of the two peaks.

  RXTE Pk4 in 
Harding et al. (2002) will be explained by the inward synchrotron
 emissions below the null charge surface, which creates
 a peak (denoted with IP3) 
 in Figure~\ref{pulse}. RXTE observations show that the phase of Pk4 
is aligned with the phase of the radio pulse. 
The present model also expect that the phase of the IP3 is close to the 
phase of  the radio peak. 
 Polarization studies for the observed radio emissions 
from the young pulsars indicate that the emission hight of the emission 
is between 1 and 10 per cent of the light cylinder radius 
(Johnston \& Wesberg 2006). 
For the inclination angle $\alpha=65^{\circ}$, because the radial distance 
to the null charge surface on the last open filed lines in the meridional 
plane is about 10 per cent of the light cylinder, the radio emission region 
will be located near or below null charge surface for the Vela pulsar. 
For example, if the radio emissions are occurred at the radial distance that 
 5 per cent of the light radius, 
 the radio pulse will be observed  at the pulse phase of $\Phi\sim 1$ 
for the observer with the viewing angle $\xi\sim 97^{\circ}$, 
 as the thick solid circles in Figure~\ref{map} show. Therefore, 
there is a possibility  that  we observe the pulse peak (IP3) from 
the inward emissions below the null charge at the phase aligned 
 the phase of the radio peak. 
Although the predicted phase-separation between IP2 and IP3 is smaller 
than the observed phase-separation of RXTE Pk3 and Pk4, 
it may be because   the real magnetic field structure around the neutron star 
will not be exactly described by the rotating dipole field,
 which was  assumed in the present paper.

The pulse peaks, whose phases are in phase with 
the two-peaks in X-ray band (RXTE Pk~1 and Pk~2-soft in Harding et al. 2005) 
and in $\gamma$-ray bands (OP1 and OP2 in Figure~\ref{pulse}), disappear 
in optical/UV bands  (Romani et al. 2005).   Furthermore, a new peak in 
optical/UV bands  is observed  at the phase between 
the phases of OP1 and of OP2  of Figure~\ref{pulse}. 
 We argue that  the two peaks related with the higher energy emissions 
disappear  because the primary particles do not contribute to the emissions 
in optical/UV bands, as Figure~\ref{pulse} (left) shows. 
The emissions from the  secondary pairs take place at higher altitude than 
the altitude of the primary emission regions. In such a case, the  phases 
of the pulse peaks of the outgoing secondary pairs 
is shifted inside the phases 
of the peaks produced by the outgoing primary particles. This may be reason 
 for the  shift of the observed first peak in optical/UV bands.

In summary of this section, we have calculated the phase averaged spectrum of 
the Vela pulsar 
with the two-dimensional electrodynamic outer gap model,  and we have 
explained the observations in optical
 to $\gamma$-ray bands. We found that both emissions from the 
outgoing and ingoing  parities are required to 
explain the observed spectrum. The present model 
predicts that  
the curvature radiation of the outgoing primary particles 
is the major emission process above 10~MeV, the outward and inward 
synchrotron emissions for the primary and secondary particles  
contribute to the emissions in X-ray bands, 
and the synchrotron radiation of the secondary particles 
explains the optical/UV emissions. 
We also calculated the pulse profile with a  three-dimensional outer-gap 
model. The present model predicts 
 the multi-peak pulse profile of RXTE observations with 
the outward and inward emissions. 

\subsection{PSR B1706-44}
\begin{figure}
\begin{center}
\includegraphics[width=7cm, height=7cm]{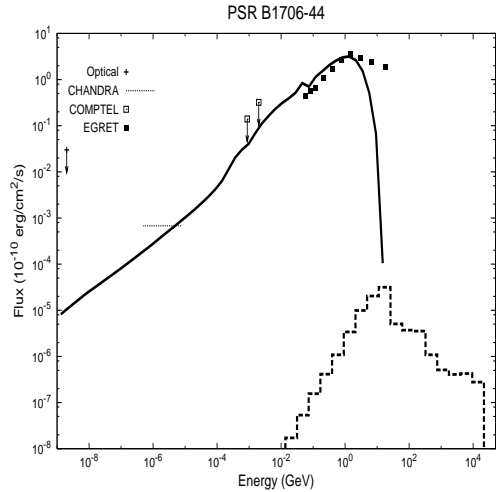}  
\caption{The spectrum of PSR B1706-44. The solid line shows the spectrum
of the total emissions, which include the synchrotron and the
curvature radiations from the outgoing and ingoing particles. The
dashed line shows the spectrum of the inverse~-Compton process of the
primary particles. The observation data are taken from Chakrabarty \&
Kaspi (1998) for optical, Gotthelf et al. (2002) for CHANDRA,
Carrami\~{n}ana et al. (1995) for COMPTEL and Thompson et al. (1996)
for EGRET.}
\label{B1706spe}
\end{center}
\end{figure}
In this section, we apply the model to PSR B1706-44.
PSR B1706-44 is the twin of the Vela pulsar. The magnetic field strength on 
the stellar surface and the spin down age of the PSR B1706-44 are similar 
to the Vela pulsar. However, the spectral behavior above the spectral 
energy peaks observed by the EGRET are different. The EGRET spectrum 
of the PSR B1706-44 extends above 10~GeV bands from the energy peak, 
$\sim 1~$GeV, with a larger spectral index than that of the Vela pulsar. 
In the X-ray bands, although CHADRA and XMM-Newton observations 
have detected the non-thermal  
emissions, the emission mechanism has not been discussed in detail.  
Because the inclination angle is not constrained well for this pulsar,
 we adopt an inclination angle  of $\alpha=65^{\circ}$ of the Vela pulsar. 
For the surface 
X-ray emissions, we apply the temperature $kT_s=0.14$ with the effective 
radius $R_{eff}\sim4  (d/2.5~\mathrm{kpc})^2$~km (Gotthelf et al. 2002).

Figure~\ref{B1706spe}
 shows the spectrum of the total emissions, which include the 
emissions for the primary and the secondary particles. The solid line 
shows the spectrum of the synchrotron and curvature radiation,  and the 
dashed line represents the spectrum of the inverse~-Compton process. 
As the solid line shows, the calculated X-ray emissions explain 
 the flux of CHANDRA observations 
but the predicted spectral index $s\sim 0.5$, 
which is defined by $I_{\nu}\propto \nu^{-s}$, does not explain 
 the index $s\sim1$ of CHANDRA observation. However,  the present model is 
consistent with   
 XMM-Newton observation of PSR B1706-44, which  indicates 
the spectral index of $s\sim 0.5$ (McGowan et al. 2004).

In $\gamma$-ray bands, the predicted spectrum explains the EGRET observations 
below 1~GeV very well. However, the observed 
spectral slope above 1~GeV is not explained by the present model, which 
predicts the exponential decrease of the flux beyond the spectral peak 
energy. The observed harder spectral index beyond the spectral peak energy
 may be related with the three-dimensional effect which due to that 
 the line of sight of the observer passes the outer gap.  
Also,  because  EGRET observed the photons 
around 10~GeV bands with a less sensitivity, 
it is important to  observe the $\gamma$-ray emissions around 10~GeV  with 
a better sensitivity.  It will be done by GLAST. 

\subsection{PSR 1951+32}
\begin{figure}
\begin{center}
\includegraphics[width=7cm, height=7cm]{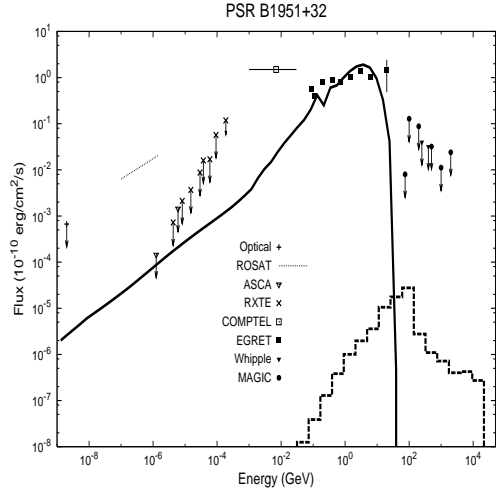}  
\caption{The spectrum of PSR B1951+32. The solid line shows the spectrum
of the total emissions, which include the synchrotron and the
curvature radiations from the outgoing and ingoing particles. The
dashed line shows the spectrum of the inverse~-Compton process of the
primary particles. The observation data are taken from Kulkarni (1988)
for optical, Safi-Harb et al.  (1995) for ROSAT,
Chang \& Guo (2000) for ASCA and RXTE, Kuiper et al. (1998) for 
COMPTEL, Ramanamurthy et al. (1995) for EGRET, and Albert et al. (2007) for
Whipple and MAGIC.}
\label{B1951spe}
\end{center}
\end{figure}
PSR B1951+32 is one of the older $\gamma$-ray pulsars with 
the spin down age, $\tau\sim1.1\times 10^5$~yr,
 and has a stellar magnetic field of 
$B_s\sim 9.7\times 10^{11}$~Gauss. The property of the X-ray emissions 
of B1951+32 is not constrained observationally, because the X-ray and soft 
$\gamma$-ray emissions is so weak.  
For example, the X-ray emission property determined by
  ROSAT observations (Safi-Harb et al. 1995) 
conflicts with 
the upper limits determined by the  RXTE and ASCA observations 
(Chang \& Ho 1997; Chang \& Guo 2000). We apply the theory to PSR B1951+32 
to show the predicted spectrum from 
optical through $\gamma$-ray bands with the present model. 

 We infer the surface X-ray emissions of PSR B1951+32 from  
 the values of the Geminga pulsar, 
which is a  $\gamma$-ray pulsar with  a similar 
spin down age, $\tau\sim 3\times 10^5$~yr, with PSR B1951+32.
 For the Geminga pulsar, 
the surface temperature is $kT_s=0.043$~keV with an effective area 
$R_{eff}\sim 13 
(d/0.12~\mathrm{kpc})^2$ (Kargaltsev et al. 2005), which means that 
the observed thermal emissions are
 most likely emitted from the bulk of the neutron star surface. For 
PSR B1951+32, therefore,  we 
adopt  the surface temperature $kT_s=0.043$~keV 
with an effective radius $R_{eff}\sim 13 (d/2.5~\mathrm{kpc})^2$.
 We use an inclination angle of $\alpha=65^{\circ}$. 

Figure~\ref{B1951spe} shows the calculated spectrum of the total
emissions, which include 
the emissions for the primary and the secondary particles.
The solid line is the spectrum of the synchrotron and the curvature 
radiation, 
and the dashed line shows the spectrum of the inverse~-Compton process. 
 As the solid line shows, the present model predicts that 
the X-ray and soft $\gamma$-ray  emissions are so week 
to detect with  the  present instruments, and the result does not
 explain the fluxes determined by ROSAT and COMPTEL.
 On the other hand, the present 
result is consistent with the upper limits determined by the  RXTE and ASCA 
observations. Above 100~MeV, the model spectrum explains 
the EGRET observations, and is consistent with the upper limits determined by 
Whipple and MAGIC observations.

\section{Discussion}
\subsection{Phase resolved-spectrum}
With the RXTE observations, 
 Harding et al. (2002) proposed the spectral index $s$,
 which is defined by $I_{\nu}\propto \nu^{-s}$, of  $s\sim 1$ 
for  the phase-resolved spectrum of the peak (Pk~2-soft in figure~1 of 
Harding et al, 2002), 
which appears between the two peaks, whose phases 
are aligned with the peak positions 
in $\gamma$-ray bands. 
As we discussed in section~\ref{pulsep}, the  preset model explains 
emission mechanism of RXTE Pk~2-soft with
 the synchrotron emissions of the ingoing 
particles, which create a peak (Figure~\ref{pulse}, IP1)
 in the pulse profile. 
 Although the spectrum in Figure~\ref{spe1} is 
calculated using two-dimensional model, we may be able to read 
 the spectral index of the phase-resolved spectra
 of the peak IP1 from Figure~\ref{spe1} 
( from the slope of the thin dashed-dotted line in the right panel), which  
predicts the spectral index $s\sim 0$ for the IP1 emissions. 
 For the reason of this discrepancy between the spectral indexes of RXTE
 and the present model, 
 we argue that 
the three-dimensional structure effects the phase-resolved spectrum, and/or
 that  RXTE observations might  not determine the 
spectral behavior in the X-ray bands very well due to a bright synchrotron 
X-ray nebula. The present model predicts the inward synchrotron
 emissions for the ingoing primary particles (Figure~\ref{spe1}, 
the thin dashed line in the left panel) contribute to the spectrum 
in hard X-ray and
 soft $\gamma$-ray bands. Therefore it is important 
to measure the phase-resolved spectra and the pulse profiles of  
the Vela pulsar in the hard X-ray and soft $\gamma$-ray bands to see 
how  the phase-resolved spectra and the pulse profiles evolve.

\subsection{Correlation between radio emission and the outer gap emission} 
Recently, the correlation between the arrival time of the radio
 emissions and the shape of the non-thermal 
 X-ray pulse profile of the Vela pulsar was  discovered (Lommen et al. 2007). 
Although the radio emission mechanism has not been understood well,
 the polarization measurement indicates a correlation between 
the magnetic field geometry of the 
 polar cap region and the swing of the position angle of the polarization 
of the radio emissions. Therefore, the polar cap accelerator model 
as the origin of the radio emissions from the pulsar has been widely accepted. 
 A  magnetospheric 
model having both  the polar cap 
accelerator and the outer gap accelerator has been proposed
 upon request to balance between the energy loss rate and the 
angular momentum loss rate, which is equal to the energy loss rate divided by 
the angular velocity of the star, of the global magnetosphere
 (Shibata 1991; 1995).
Although  the polar cap  accelerator and the outer gap 
accelerator can not exist on the same magnetic field lines because 
of the different current directions in the polar cap and outer gap
 accelerators, the polar cap 
accelerator affects the outer gap accelerator to obtain 
the torque balance. According to this model, therefore,  even though 
the  non-thermal X-rays originate from the outer gap accelerator, some 
correlations between X-ray emissions 
and the radio emission are expected. 
Furthermore, the present  model also predicts that
 the outer gap accelerator  affects the polar cap accelerator with the inward 
emissions, which pass near the stellar surface.  
The  inward emissions will affect the circumstance 
around polar cap region through  the pair-creation process 
and/or the scattering process.

\subsection{Inward emissions from the Crab pulsar}
\label{crab}
Unlike the Vela pulsar, the Crab pulsar has only two peaks in the pule profile 
in  optical through $\gamma$-ray bands. 
This indicates that the observed emissions of the Crab pulsar are dominated by 
the outward emissions  in whole energy bands. 
For the Crab pulsar, 
the  photons above 1~GeV  emitted 
via the curvature radiation in the outer gap   are converted into 
the pairs outside the gap by the pair-creation process by 
 the magnetospheric X-ray photons emitted by the secondary pairs. 
Outer gap model predicts the synchrotron and the inverse~-Compton processes 
of the secondary pairs explain the observed emissions in optical through
$\gamma$-ray bands. As we discussed in section~\ref{spectrum} 
(Figure~\ref{spe1},right), the outward 
synchrotron radiation by the outgoing secondary pairs created by 
the magnetospheric X-rays is much brighter than the inward synchrotron 
radiation, because the outgoing $\gamma$-rays emitted by the 
curvature radiation is much more than the ingoing $\gamma$-rays.  
For the Crab pulsar, therefore, we can expect that the outgoing secondary 
pairs produce  the observed spectrum with 
the synchrotron 
and the inverse~-Compton process.  Because the flux of the outward emissions 
of the secondary particles 
is much (say one or two order) larger than that of the inward emissions, 
the pulse profiles have strong two peaks in 
a single period in optical through $\gamma$-ray bands for the Crab pulsar.  
Although we expect contribution of the inward emissions at off-pulse phase, 
 it must be difficult to separate the tiny flux of the inward emissions 
from off-pulse outward emissions inside null charge surface and/or 
from the strong background  emissions from the synchrotron nebula.  

\section*{Acknowledgments}

The authors appreciate fruitful discussion with K.S. Cheng, 
K.Hirotani and R.Taam. This work was supported by the Theoretical
Institute for Advanced Research in Astrophysics (TIARA) operated under
Academia Sinica and National Science Council Excellence Projects
program in Taiwan administered through grant number NSC 96-2752-M-007-001-PAE.

\label{lastpage}

\end{document}